\documentclass[%
 preprint, 
 amsmath,amssymb,
 aps, physrev,
]{revtex4-2}

\numberwithin{equation}{section}

\usepackage{babel}
\usepackage{graphicx}
\usepackage{dcolumn}
\usepackage{bm}
\usepackage[compat=1.1.0]{tikz-feynman}
\usepackage{feynmp-auto} 
\usepackage{tikz}

\begin{document}


\title{\textbf{The Modification of Feynman Diagrams in Curved Space-Time} 
}%

\author{Benliang Li\\libenliang732@gmail.com}
\affiliation{
 School of Mathematics and Physics, University of South China, 421001 Hengyang, China}%
 
\affiliation{\\
 Dipartimento di Fisica e Astronomia, Universita di Bologna, 40126 Bologna, Italy
}%

\date{\today}

\begin{abstract}
This paper explores the behavior of quantum particles in weak gravitational fields. We examine scalar and spinor particles, showing that these quantum particles in weak gravitational fields follow geodesic trajectories, aligning with classical expectations. Further, we explore the impact of gravitational fields on Yukawa interaction $\lambda \bar{\Psi} \Phi \Psi$, revealing that Feynman diagrams are modified due to the curvature, affecting propagators and interaction dynamics. The presence of spatially-dependent factors in propagators underscores the localized nature of gravitational effects on particle interactions. 
\end{abstract}

\maketitle




\section{Introduction}\label{Sec: one}

The interplay between quantum field theory (QFT) and general relativity remains one of the most intriguing areas in modern theoretical physics. As researchers continue to probe the effects of curvature on quantum fields, the need to understand particle behavior in curved space-time becomes ever more critical. 
 
The foundational understanding of particles within the context of curved spacetime has been a pivotal aspect of modern theoretical physics since the development of Einstein's theory of General Relativity. The concept of particles, fundamental to the fabric of matter and interactions, becomes inherently complex and elusive when applied to the curved spacetime framework. While classical particle concepts are well-defined in flat spacetime, their extension to curved spacetime presents intriguing challenges and necessitates a deeper reevaluation. In the standard formulation of quantum field theory in curved space-time (QFT-CS), particles are typically described as eigenfunctions of the field equation, with the Feynman propagator defined as the inverse of the second-order differential operator associated with the field equation in curved spacetime \cite{birrell1984quantum, parker2009quantum, bunch1979feynman, parker1985new, markkanen2013simple}. In this paper, we propose a departure from this conventional perspective, arguing that identifying particles as eigenfunctions of field equations in curved spacetime may be a potentially misleading representation of their true nature. We introduce an alternative approach to QFT-CS that addresses these conceptual challenges. 

The structure of this paper is organized as follows: In Section 2, we perform the quantization of fields within a curved spacetime framework using the Local Minkowski Coordinate (LMC) approach. We also reproduce the classical geodesic path by examining the conservation of energy-momentum principles in the context of quantum fields, followed by a discussion on the misconceptions surrounding particles. Sections 3 and 4 build upon this foundation by exploring case studies on the behavior of Klein-Gordon (KG) and Dirac particles within the Weak Schwarzschild Space-Time (WSST) framework. Section 5 addresses the modifications to Feynman diagrams in curved space-time, specifically within the WSST framework. By analyzing Yukawa $\lambda \bar{\Psi} \Phi \Psi$ theory, we demonstrate how the gravitational field alters propagators and scattering probabilities. The inclusion of spatially-dependent factors and the impact of the metric tensor in Feynman rules underscore the importance of incorporating curvature effects into theoretical predictions. This adaptation is crucial for accurately describing particle interactions in realistic gravitational settings.

In this paper, Greek characters (e.g., $\mu$, $\nu$) are used as indices for global coordinates established in curved space-time, while Latin characters (e.g., $a$, $b$, $i$, $j$) represent indices for LMC. The Minkowski metric follows the convention $\eta_{ab}=diag(-1,1,1,1)$, and the symbol $a\in(0,1,2,3)$ is applied consistently throughout this paper. Throughout this paper, we adopt natural units with $\hbar=c=1$.

\section{Conception of Quantum Particles and Geodesic Motion in Weak Gravitational Fields}

QFT is a powerful framework that describes the behavior of elementary particles and their interactions, grounded in the principles of quantum mechanics and special relativity. Although QFT is traditionally formulated in flat Minkowski space-time, many astrophysical phenomena and cosmological scenarios necessitate an extension of QFT to curved space-time. In this extended context, it becomes essential to define a coordinate system that locally approximates the Minkowski metric. These coordinates, known as "Local Minkowski Coordinates" (LMC), are instrumental in studying quantum fields in the presence of gravity. LMC allow us to analyze quantum field dynamics locally, providing a familiar Minkowski-like framework within small regions of curved space-time.

In this section, we will explore the significance of LMC in QFT-CS and their role in facilitating the study of particle interactions and quantum phenomena within the complex landscape of curved space-time. By addressing the challenges posed by curved space-time and the necessity of a local Minkowski approximation, we lay the groundwork for a comprehensive understanding of QFT-CS. Our investigation will focus on key areas of interest, including geodesic motion, energy-momentum tensor (EMT), and the intricate connection between LMC and global coordinates in curved space-time.

Considering the action in curved space-time, we have:

\begin{equation}
S = \frac{1}{16\pi G} S_H + S_M,
\end{equation} 
where $S_H = \int \sqrt{-g}Rd^{4}x$ is known as the Hilbert action, with $R$ denoting the Ricci scalar and $g$ being the determinant of the metric tensor. $S_M$ represents the action of the matter fields. By varying Eq. (2.1) with respect to the metric tensor $g_{\mu \nu}$, we obtain the Einstein field equations. The EMT of the matter fields is given by:
\begin{equation}
T^{\mu \nu} = -2\frac{1}{\sqrt{-g}} \frac{\delta S_M}{\delta g_{\mu \nu}}.
\end{equation} 

The conservation of the EMT is expressed as:
\begin{equation}
\nabla_{\mu}T^{\mu \nu} = 0
\end{equation} 
where $\nabla_{\mu}$ represents the covariant derivatives. This equation reflects the conservation laws in curved space-time, analogous to how the divergence of the EMT vanishes in flat space-time. To investigate the motion of a quantum particle in curved space-time, we utilize Eq. (2.3), with $T^{\mu \nu}$ determined by the relevant quantum fields. This approach allows us to explore the dynamics and interactions of particles within a curved background, providing insights into the behavior of quantum fields under the influence of gravity. 

Consider the action of the KG field in curved space-time:
\begin{equation}
S_{KG} = \int \left(-\frac{1}{2}g^{\mu \nu}\nabla_{\mu}\phi \nabla_{\nu}\phi-\frac{1}{2}m^2 \phi^2\right)\sqrt{-g}d^{4}x.
\end{equation} 
Varying  $S_{KG}$ with respect to the field  $\phi$  yields the field equation in curved space-time: 
\begin{equation}
(g^{\mu \nu}\nabla_{\mu}\nabla_{\nu}-m^2)\phi = 0.
\end{equation} 
This can be rewritten in the form: 
\begin{equation}
\frac{1}{\sqrt{-g}}\partial_{\mu}(\sqrt{-g}g^{\mu \nu}\partial_{\nu}\phi) = m^2\phi,
\end{equation} 
which facilitates solving the field equation. The general solution can be expressed as:
\begin{equation}
\phi_{k}(t,\vec{x}) = B_k(t,x)e^{-i \int k_a e^{a}_{\mu} dx^{\mu}}
\end{equation} 
where $B_k(t,x)$ is a real function to be determined. $k_a$ presents the wave vector components, and $e_{\mu}^{a}$ are the components of the tetrad (vierbein) field that relate LMC to the curved space-time coordinates. This form of the solution, commonly referred to as the Wentzel-Kramers-Brillouin (WKB) approximation, elucidates the behavior of the field under the combined effects of the mass term and the curvature of spacetime \cite{alsing2001phase, oancea2020gravitational, maniccia2023qft, hammad2024curved}.  In this paper, we analyze the dynamics of a free quantum particle propagating in a weak gravitational field (WGF), where the perturbations in the metric and the variations in $B_k(t,x)$ and $k_a$ are assumed to be small compared to the particle’s dominant energy scale. This assumption allows us to retain the WKB approximation to first order, offering valuable insights into the quantum behavior of the particle in curved space-time. 

The integration in Eq. (2.7) is performed along a specific path, which will later be identified as the geodesic path. It is important to note that the components $k_a$ are functions of the variable $x$ when integrating over $dx$ in Eq. (2.7). These components can be determined from the conservation of the EMT.


In WGF and substituting Eq. (2.7) into Eq. (2.6), we obtain: 
\begin{equation}
\eta^{ab}k_{a}k_b+m^2=0,
\end{equation} 
which is the dispersion relation for the KG field in the presence of WGF. Consequently, the quantized KG field in WGF can be expressed as: 
\begin{equation}
\begin{split}
\Phi(t,\vec{x}) = \int \frac{d^{3}{k_i}}{(2\pi)^3}\frac{1}{\sqrt{2k_0}}[a_{k_i} \phi_{k}(t,\vec{x})+a_{k_i}^{+} \phi_{k}^{\ast}(t,\vec{x})]
\end{split}
\end{equation} 
where $a^{+}_{k_i}$  and $a_{k_i}$ are the creation and annihilation operators, respectively, for a KG particle with momentum vector $k_i$ measured in a LMC system. These operators satisfy the commutation relation: 
\begin{equation}
\begin{split}
[a_{k_i},a_{k_j}^{+}] = (2\pi)^3\delta^3(k_i-k_j)\delta_{ij}
\end{split}
\end{equation} 
It's important to note that since LMCs in WGF form a continuum of coordinates defined at each space-time point, the value of the momentum $k_i$ may vary depending on the LMC in which it is measured. Therefore, the commutation relation given by Eq. (2.10) must be evaluated within the same LMC, ensuring consistency in the definitions of momentum and the corresponding operators. 

By substituting Eq. (2.7) into Eq. (2.6), we can observe that Eq. (2.7) also represents the wave function of a free KG particle propagating in curved space-time, analogous to a plane wave described by $e^{-i(\omega t-kx)}$ in global Minkowski space-time. The key difference is that in curved space-time, the probability density distribution of a free particle, given by $B^{2}_k(t,x)$, is distorted by the presence of gravity, whereas in global Minkowski space-time, the probability density distribution remains uniform for free particles. In other words, gravity not only affects the particle's energy-momentum, expressed by $k_{a}e_{\mu}^{a}$  in the phase factor $e^{-i \int k_a e^{a}_{\mu} dx^{\mu}}$, but also deforms its probability density distribution in space-time. This highlights the influence of gravitational fields on both the kinematic properties and spatial behavior of quantum particles, revealing the complex interplay between QFT and general relativity.

It is important to note that the wave with vector $k_a$ measured in LMC or with vector $k_ae^{a}_{\mu}$ measured in the global coordinates $x$ represents the same wave as described by Eq. (2.7), but measured in two different coordinate systems. When changing coordinates from $(t,\vec{x})$ to a newly defined coordinate system $(t',\vec{x}')$, the quantized KG field expressed in the new coordinates $(t',\vec{x}')$ can be written as: 

\begin{equation}
\begin{split}
\Phi^{'}(t',\vec{x}') = \int \frac{d^{3}{k_i}}{(2\pi)^3}\frac{1}{\sqrt{2k_0}}[a_{k_i} \phi^{'}_{k}(t',\vec{x}')+a_{k_i}^{+} \phi^{'\ast}_{k}(t',\vec{x}')]
\end{split}
\end{equation} 
where $\phi'_{k}(t',\vec{x}')$ is given by
\begin{equation}
\begin{split}
\phi'_{k}(t',\vec{x}') = B'_k(t',\vec{x}')e^{-i \int k_a e^{a}_{\mu'} dx'^{\mu}}
\end{split}
\end{equation} 
with $B'_k(t',\vec{x}') = B_k(t,\vec{x})$ and $e^{-i \int k_a e^{a}_{\mu'} dx'^{\mu}} = e^{-i \int k_a e^{a}_{\mu} dx^{\mu}}$ at the corresponding transformed space-time points. This demonstrates that the wave function remains a scalar under coordinate transformations. 

It is also worth noting that the creation ($a_{k_i}^{+}$) and annihilation ($a_{k_i}$) operators remain unchanged in the new coordinate system $(t',\vec{x}')$ because they are defined within the LMC. In other words, $a_{k_i}^{+}$ creates the same particle, represented by $\phi'_{k}(t',\vec{x}')$ in the coordinates $(t',\vec{x}')$ or by $\phi_{k}(t,\vec{x})$  in the coordinates $(t,\vec{x})$, from the same vacuum state. 
Consequently, the field operators $\Phi^{'}(t',\vec{x}')$ and $\Phi(t,\vec{x})$, when acting on the vacuum, create the same particle but are expressed in two different coordinate systems. This consistency underscores the coordinate-independent nature of the quantized field, highlighting the scalar property of the wave function across different coordinate frames. 

The EMT of the quantized KG field is given by:
\begin{equation}
\begin{split}
T_{\mu \nu} = \nabla_{\mu}\Phi \nabla_{\nu}\Phi-\frac{1}{2}g_{\mu \nu}(g^{\rho \sigma}\nabla_{\rho}\Phi \nabla_{\sigma}\Phi+m^2 \Phi^2)
\end{split}
\end{equation} 
in which $\Phi$ is given by Eq. (2.9). For a single KG particle moving in curved space-time, the expectation value of the EMT is expressed as: 
\begin{equation}
\begin{split}
\langle T_{\mu \nu} \rangle_k = \langle k_j \mid \nabla_{\mu}\Phi \nabla_{\nu}\Phi \mid k_j\rangle-\frac{1}{2}g_{\mu \nu}\langle k_j \mid (g^{\rho \sigma} \nabla_{\rho}\Phi \nabla_{\sigma}\Phi +m^2 \Phi^2)\mid k_j\rangle
\end{split}
\end{equation} 
where $\mid k_j\rangle \equiv \sqrt{2k_0} a_{{k_j}}^{+} \mid 0 \rangle$ represents the single particle's state with wave-vector $k_j$ measured in LMC. Plug Eq. (2.9) into Eq. (2.14), we obtain
\begin{equation}
\begin{split}
\langle T_{\mu \nu} \rangle_k = 2B_k^{2}k_{\mu}k_{\nu} +(const.)_{\mu\nu},
\end{split}
\end{equation} 
where the term $const.$ represents infinite constants that act similarly to the expectation value of the EMT of the vacuum. These constants can be discarded as they do not influence our results. The energy-momentum vector of a single particle, measured in the global coordinate system, is denoted by $k_{\mu}\equiv k_{a} e^{a}_{\mu}$. This vector satisfies the relation:
\begin{equation}
g^{\mu \nu}k_{\mu}k_{\nu}+m^2=0.
\end{equation} 
which is the mass-shell condition for the particle in curved space-time. 

The current of the single KG particle is given by:
\begin{equation}
\begin{split}
j_{k}^{\mu}=\frac{i}{2m}(\phi_{k}^{+}\nabla^{\mu}\phi_{k}-\phi_{k}\nabla^{\mu}\phi_{k}^{+})=\frac{B_{k}^{2}k^{\mu}}{m}
\end{split}
\end{equation} 
where the wave function of the single particle $\phi_k$ is given by Eq. (2.7). The continuity equation,
\begin{equation}
\nabla_{\mu}j_{k}^{\mu}=0
\end{equation} 
implies $\nabla_{\mu}(B_{k}^{2}k^{\mu})=0$. Consequently, the conservation of the EMT, $\nabla_{\mu}\langle T^{\mu \nu} \rangle=0$, leads to the result:
\begin{equation}
\begin{split}
B_{k}^{2}k^{\mu}\nabla_{\mu} k^{\nu}=0\Longrightarrow k^{\mu}\nabla_{\mu} k^{\nu}=0.
\end{split}
\end{equation} 
Equation (2.19) takes the same mathematical form as the classical geodesic equation. Therefore, requiring the conservation of the EMT for a KG particle moving in a WGF reproduces the geodesic equation, showing that the particle follows a geodesic path. 

\subsection {comments on geodesic path and the notion of particles}

The geodesic equation can be derived from QFT-CS through various methods. The simplest approach involves applying the covariant derivative  $\nabla_{\mu}$  to Eq. (2.16) alongside the following condition:
\begin{equation}
\nabla_{\nu}k_\mu =\nabla_{\mu}k_\nu
\end{equation} 
However, we have demonstrated an alternative method in which the wave solution presented in Eq. (2.7) does not require satisfying Eq. (2.20) to reproduce the geodesic equation. Specifically, the geodesic path can be derived solely from the conservation of the energy-momentum, without the need for the symmetry condition on the covariant derivatives of $k_\mu$. This approach provides a more flexible pathway to linking quantum field solutions with classical geodesic motion in curved spacetime.

The wave solution given by Eq. (2.7) is typically interpreted as the first-order WKB approximation. However, it is important to emphasize that these solutions are not eigenfunctions of the field equation. For instance, in WSST, we have: 

\begin{equation}
\frac{1}{\sqrt{-g}}\partial_{\mu}(\sqrt{-g}g^{\mu \nu}\partial_{\nu}\phi_{k}) \approx g^{\mu \nu}k_{\mu}k_{\nu}\phi_k
\end{equation} 
where $\phi_k$ is given by Eq. (2.7). Consequently,  $\phi_k$  is not an eigen-function since the momentum $k_a\equiv e^{\mu}_{a}k_{\mu}$ are not constant factors (or eigenvalues) in Schwarzschild space-time.

The interpretation of particles in curved space-time is a subject of ongoing debate, with a widely held view that particles are mathematically represented as eigenfunctions of the field equations. However, within the context of Schwarzschild spacetime, eigenfunctions of the form $R_{\omega,l}(r)Y_{l,m}(\theta,\varphi)e^{i\omega{}t}$ \cite{elizalde1988exact, qin2012exact, lehn2018klein} , where $\omega$, $l$ and $m$ are the eigenvalues of the second-order differential operator associated with the field equation, fail to approximate the geodesic equation as described by Eq. (2.19). Furthermore, they do not asymptotically approach ordinary plane waves as the space-time becomes Minkowski-like at spatial infinity. In contrast, the geodesic equation can only be derived from the solution Eq. (2.7), indicating that these solutions represent actual physical particles in curved spacetime. 
Indeed, if one were to consider the eigenfunction $R_{\omega,l}(r)Y_{l,m}(\theta,\varphi)e^{i\omega{}t}$ as representing a physical particle in Schwarzschild space-time, the corresponding free Hamiltonian would be quantized as $H_0=\sum_{\omega,l,m}\omega a_{l,m,\omega}^{+}a_{l,m,\omega}$ where $a_{l,m,\omega}^{+}$ denotes the creation operator associated with the particle described by $R_{\omega,l}(r)Y_{l,m}(\theta,\varphi)e^{i\omega{}t}$. The evolution of the particle in free Schwarzschild space-time is then governed by 
\begin{equation}
e^{-iH_{0}t}a_{l,m,\omega}^{+}\mid 0 \rangle=e^{-i\omega t}a_{l,m,\omega}^{+}\mid 0 \rangle
\end{equation} 
where $\mid 0 \rangle$  represents the vacuum state. From Eq. (2.22), it is clear that the propagation of the free particle in Schwarzschild spacetime remains invariant, contradicting experimental observations that indicate particle motion should follow geodesic paths. Therefore, interpreting these eigenfunctions as physical particles in curved spacetime presents significant inconsistencies. 
To enhance our understanding, we can draw an analogy with classical physics. In classical physics, the EMT is expressed as $T_{\mu \nu}\equiv \rho u_{\mu}u_{\nu}$, where $\rho$  and $u$ represent the density and four-velocity of classical particles, respectively. Analogously, $B^{2}_{k}$  and $k_{\mu}$ in Eq. (2.15) represent the probability density distribution and momentum of quantum particles in curved space-time. This analogy breaks down if we interpret eigenfunctions like  $R_{\omega,l}(r)Y_{l,m}(\theta,\varphi)e^{i\omega{}t}$ instead of Eq. (2.7) as physical particles. Notably, these two solutions are obtained in different coordinate systems: the former in spherical coordinates and the latter in Cartesian coordinates. It is crucial to exercise caution when performing coordinate transformations in curved spacetime, as the resulting solution obtained in the transformed coordinates may no longer correspond to the original physical problem \cite{lin2024revisit}. This underscores the necessity for re-evaluating how we interpret particles within the framework of QFT-CS to ensure that their behavior aligns with the expected geodesic motion in WGF. 

In summary, to ensure consistency with the established principles of particle motion in curved spacetime, it is imperative that any interpretation of quantum particles accurately reflects geodesic behavior in WGF. We showed that the wave solutions given by Eq. (2.7), which are typically seen as first-order WKB approximations, offer a more accurate representation of physical particles in curved spacetime than conventional eigenfunctions. These wave solutions align with the expected geodesic motion of particles, unlike the commonly used eigenfunctions, which fail to reproduce the geodesic equation and do not match the asymptotic behavior of plane waves in a Minkowski-like spacetime at spatial infinity. 

In the following sections, we will continue to apply our quantization procedure in LMC to some cases and explore particle interactions in curved space-times.

\section{Case Study 1: Behavior of KG Particles in Weak Schwarzschild Space-Time (WSST)}

In Schwarzschild space-time, the metric in spherical coordinates is given by:

\begin{equation}
\begin{split}
ds^2=-(1-\frac{r_s}{r})dt^{2}+(1-\frac{r_s}{r})^{-1}dr^2+r^2d\theta^2+r^2\sin^2\theta d\varphi^2
\end{split}
\end{equation} 
where $r_s$ is the Schwarzschild radius. Transforming Eq. (3.1) into a "Cartesian-like" coordinate, the metric becomes:

\begin{equation}
g_{\mu\nu}(t,x,y,z)=
\left(
\begin{array}{cccc}
-\frac{r-{r_s}}{r}& 0& 0& 0 \\
0 & 1+\frac{r_s x^2}{(r-r_s) r^2} & \frac{r_s x y}{(r-r_s) r^2} & \frac{r_s x z}{(r-r_s) r^2}\\
0 &  \frac{r_s x y}{(r-r_s) r^2} & 1+\frac{r_s y^2}{(r-r_s) r^2} & \frac{r_s y z}{(r-r_s) r^2}\\
0 & \frac{r_s x z}{(r-r_s) r^2} & \frac{r_s y z}{(r-r_s) r^2} & 1+\frac{r_s z^2}{(r-r_s) r^2}\\
\end{array}
\right)
\end{equation} 
where the relationships between the spherical and "Cartesian-like" coordinate systems are given by $r=\sqrt{x^2+y^2+z^2}$, $\cos \theta= z/r$, and $\tan\varphi=y/x $. Note that  $\sqrt{-g}=1$ for the metric given by Eq. (3.2). The "Cartesian-like" coordinates are introduced in analogy to Cartesian coordinates in Minkowski spacetime, allowing for more convenient computations, particularly for comparing Feynman diagrams in Schwarzschild and Minkowski spacetimes. In standard QFT, plane waves are typically utilized as solutions to free-field equations (such as the Klein-Gordon and Dirac equations) and serve to represent incoming and outgoing particles in S-matrix calculations. If spherical coordinates were used, the solution would take the form of a spherical wave rather than a plane wave, complicating direct comparisons. Consequently, the adoption of "Cartesian-like" coordinates here aligns with standard QFT practices in flat spacetime. 

Additionally, if we solve Eq. (2.6) within the metric defined in a LMC system, as seen in previous works such as Refs. \cite{mandal2021local, perche2022localized}, the resulting solution would be limited to that LMC. It would not remain valid when transformed into a different LMC. In contrast, a solution derived in a global coordinate system applies universally across the entire Schwarzschild spacetime. This allows us to compare S-matrix calculations across different LMCs, which is the primary objective of this work. 

The vierbein $e_{\mu}^{a}$  satisfies the relation $g_{\mu\nu}=e^{a}_{\mu}e^{b}_{\nu}\eta_{ab}$, it can be expressed as: 
\begin{equation}
e_{\mu}^{a}=
\left(
\begin{array}{cccc}
\frac{1}{f_{r_s}+1}& 0& 0& 0\\
0& 1+\frac{x^{2}}{r^{2}}f_{r_s} & \frac{xy}{r^{2}}f_{r_s} & \frac{xz}{r^{2}}f_{r_s}\\
0 & \frac{xy}{r^{2}}f_{r_s} & 1+\frac{y^{2}}{r^{2}}f_{r_s}& \frac{yz}{r^{2}}f_{r_s}\\
0 & \frac{xz}{r^{2}}f_{r_s} &  \frac{yz}{r^{2}}f_{r_s}&  1+\frac{z^{2}}{r^{2}}f_{r_s}

\end{array}
\right)
\end{equation} 
in which $f_{r_s}\equiv \sqrt{r/(r-r_s)}-1$. Note that $e_{\mu}^{a}$ is a symmetric matrix, for instance, $e_{z}^{2}=e_{y}^{3}=\frac{yz}{r^{2}}f_{r_s}$. The inverse matrix of $e^{a}_{\mu}$  can be given as
\begin{equation}
e^{\mu}_{a}=
\left(
\begin{array}{cccc}
f_{r_s}+1& 0& 0& 0  \\
0& 1-\frac{x^{2}}{r^{2}}\frac{f_{r_s}}{f_{r_s}+1} & -\frac{xy}{r^{2}}\frac{f_{r_s}}{f_{r_s}+1} & -\frac{xz}{r^{2}}\frac{f_{r_s}}{f_{r_s}+1}\\
0 & -\frac{xy}{r^{2}}\frac{f_{r_s}}{f_{r_s}+1} & 1-\frac{y^{2}}{r^{2}}\frac{f_{r_s}}{f_{r_s}+1} & -\frac{yz}{r^{2}}\frac{f_{r_s}}{f_{r_s}+1}\\
0 & -\frac{xz}{r^{2}}\frac{f_{r_s}}{f_{r_s}+1} &  -\frac{yz}{r^{2}}\frac{f_{r_s}}{f_{r_s}+1} &  1-\frac{z^{2}}{r^{2}}\frac{f_{r_s}}{f_{r_s}+1}

\end{array}
\right)
\end{equation} 
For a particle moving outside the Schwarzschild radius with the metric given by Eq. (3.2), the largest value of $\partial_{\rho}g_{\mu \nu}$ is approximately $\frac{r_s}{r(r-r_s)}$. This value only becomes significantly large when $r\approx r_s$, indicating that the particle is very close to the Schwarzschild radius where WGF no longer holds. 
Considering an electron on the surface of the Sun with a wavelength equal to the Bohr radius, we have $r_s \approx 3$ km as the Schwarzschild radius and $R_{\text{sun}} \approx 7\times 10^{5}$ km as the radius of the Sun. In this scenario, we can verify that  $k_t \approx 10^{11} \text{m}^{-1} \gg 6\times 10^{-15} \text{m}^{-1} \approx \frac{r_s}{R_{\text{sun}}(R_{\text{sun}}-r_s)}$. Therefore, the WGF is robustly maintained within our solar system, and the field equation with the metric Eq. (3.2) yields the solution given by Eq. (2.7).

In the case where  $x=y=0$ and $k_x=k_y=0$, the KG particle propagates along the $z$-direction with $z=r$, and the particle's two-dimensional wave function can be expressed as:
\begin{equation}
\begin{split}
\phi_{k}(t,z)= B_k(z)e^{-i (\omega t+\int k_{z}dz) }
\end{split}
\end{equation} 
Applying the continuity equation, we have:
\begin{equation}
\begin{split}
0=\nabla_{\mu}j_{k}^{\mu}=\nabla_{\mu}(B_{k}^{2}k^{\mu})\Rightarrow \partial_{t}(B_{k}^{2}k^{t})+\partial_{z}(B_{k}^{2}k^{z})=\partial_{z}(B_{k}^{2}k^{z})=0
\end{split}
\end{equation} 

Consequently, we obtain
\begin{equation}
\begin{split}
B_{k}=\sqrt{\frac{k^{z}(r_0)}{k^{z}}}
\end{split}
\end{equation} 
where $k^{z}=\sqrt{\omega^2-m^2(1-\frac{r_{s}}{r})}$. Here, $k^{z}(r_0)\equiv \sqrt{\omega^2-m^2(1-\frac{r_{s}}{r_0})}$ ensures that $B_{k}=1$ when $r=r_0$, where $r_0$ is chosen as a reference point. 
 

The function $B_k$, representing the particle's probability density distribution, is position-dependent in WSST. This suggests that the size of the particle may be compressed or stretched due to the gravitational field, illustrating the influence of spacetime curvature on particle dynamics.

\section{Case study 2: Dirac particle in WSST}
Dirac equation in curved space-time is given by
\begin{equation}
\begin{split}
[i\gamma^a e_{a}^{\mu} (\partial_{\mu}+\frac{1}{8} \omega_{\mu}^{cd}[\gamma_{c},\gamma_{d}])-m]\psi(t,\vec{x})=0
\end{split}
\end{equation} 
where the spin connection $\omega_{\mu}^{ab}$ is defined as $\omega_{\mu}^{ab}=e_{\nu}^{a}e^{\sigma b}\Gamma_{\mu\sigma}^{\nu}- e^{\nu b}\partial_{\mu}e_{\nu}^{a}$. The $\gamma^{a}$ are Dirac matrices in Minkowski space-time with chiral representation: $\gamma^{0}= \left(\begin{array}{cc} 0& 1  \\
1 & 0
\end{array}
\right)$ and $\gamma^{i}= \left(\begin{array}{cc} 0& \sigma ^i  \\
-\sigma ^i & 0
\end{array}
\right)$,  $\sigma ^i $ are three Pauli matries.  In the WSST, such as on the Earth's surface, the quantum particle is centered along the $z$-axis and has a characteristic size on the order of the Bohr radius, i.e., $L_x=L_y\approx a_0=5.29\times 10^{-11} m$, indicating that the particle's wavelength is on the order of $ 10^{-11}m$. Thus, the off-diagonal elements with factors $x/r$ or $y/r$ in Eq. (3.2) have an order of magnitude of  $10^{-17}$, where $r\approx 6\times 10^{6}m$ is approximately the Earth's radius. As a result, for a quantum particle in WSST, neglecting terms with factors $x/r$ ($y/r$) and higher-order terms, only six elements of $\Gamma_{\mu\sigma}^{\nu}$ contribute. These terms are: 
\begin{equation}
\begin{split}
\Gamma^{t}_{tz}=\Gamma^{t}_{zt}=\frac{1}{2}g^{tt}\partial_{z}g_{tt},~~
\Gamma^{z}_{tt}=-\frac{1}{2}g^{zz}\partial_{z}g_{tt},~~
\Gamma^{z}_{xx}=g^{zz}\partial_{x}g_{xz},~~
\Gamma^{z}_{yy}=g^{zz}\partial_{y}g_{yz},~~
\Gamma^{z}_{zz}=\frac{1}{2}g^{zz}\partial_{z}g_{zz}
\end{split}
\end{equation} 
Similarly, for the partial derivatives of the vierbein, only six terms survive, given by:

\begin{equation}
\begin{split}
\partial_{z}e_{t}^{0}=\frac{r_{s}\sqrt{z}}{2z^{2}\sqrt{z-r_s}},~~~
\partial_{z}e_{z}^{3}=-\frac{r_{s}\sqrt{z-r_s}}{2(z-r_s)^{2}\sqrt{z}},~~~
\partial_{x}e_{x}^{3}=\partial_{x}e_{z}^{1}=\partial_{y}e_{y}^{3}=\partial_{y}e_{z}^{2}=\frac{\sqrt{z/(z-r_s)}-1}{z}
\end{split}
\end{equation} 
Note that at this stage, $r$ can be replaced with $z$ since $x/z=y/z\approx 10^{-17}$. Consequently, for the spin connection $\omega_{\mu}^{ab}$, only six terms contribute, which are: 

\begin{equation}
\begin{split}
\omega_{t}^{03}=-\omega_{t}^{30}=\frac{r_s}{2z^2},~~~
\omega_{x}^{13}=\omega_{y}^{23}=-\omega_{x}^{31}=-\omega_{y}^{32}=\frac{\sqrt{(z-r_s)/z}-1}{z}
\end{split}
\end{equation} 
For a typical electron on the Earth's surface, with a wavelength of the same magnitude as the Bohr radius, i.e., $\lambda\approx a_0$, we have $\partial_z\approx k_z \approx 10^{11} m^{-1}$, while  $ \omega_{x}^{31} =\omega_{y}^{32} \approx \omega_{t}^{03}\approx 10^{-16} m^{-1} $. Therefore, the contribution from the spin connection terms is negligible in our planet.
However, near a black hole, the spin connection terms, which represent the local curvature of space-time inside the atom, cannot be neglected. Investigations into the energy level shifts caused by these terms can be found in Ref.~\cite{parker1980one, parker1982gravitational}. Thus, Eq. (4.1) simplifies to:

\begin{equation} (i\gamma^\mu \partial_{\mu}-m)\psi(t,\vec{x})=0. \end{equation} 
The solution can be given as
\begin{equation}
\psi^{s}_{k}(t,\vec{x})=B_k(\vec{x})u^{s}_{k_i} e^{-i \int k_{a} e^{a}_{\mu} dx^{\mu}},
\end{equation} 
with a real function $B_k$ and
\begin{equation}
u^{s}_{k_i}=\left(
\begin{array}{cc}
\sqrt{k_0 +k_i \sigma ^i}\epsilon^s  \\
\sqrt{k_0 -k_i \sigma ^i}\epsilon^s
\end{array}
\right)
\end{equation} 
in which $\epsilon^s$ satisfies $\epsilon^{s+}\epsilon^{s'}=\delta^{ss'}$, and it is a two-component normalized vector representing the spin direction measured in LMC. Also, $k_{0}^2=k_{i}^{2}+m^2$ is the energy-momentum relation in LMC. The quantized field for electrons can be expressed as follows:
\begin{equation}
\Psi(t,\vec{x})=\int\frac{d^{3}{k_i}}{(2\pi)^3}\frac{1}{\sqrt{2k_0}}\sum_{s=1}^{2}c^{s}_{k_i} \psi^{s}_{k}(t,\vec{x})
\end{equation} 
Here, $c^{s}_{k_i}$ represents the annihilation operator of a spin-$s$ electron with $k_i$ satisfying the anti-commutation relation $[c^{s}_{k_i},c^{s'+}_{k'_j}]_{+}=(2\pi)^{3}\delta^{ss'}\delta^{3}(k_i-k'_j)\delta_{ij}$. The EMT is given by
\begin{equation} T^{\mu \nu}=\frac{i}{2}[\bar{\Psi}(\gamma^{\mu}\nabla^{\nu}+\gamma^{\nu}\nabla^{\mu})\Psi-(\nabla^{\mu}\bar{\Psi}\gamma^{\nu}+\nabla^{\nu}\bar{\Psi}\gamma^{\mu})\Psi] \end{equation} 
Considering WGF with $\nabla_{\mu}=\partial_{\mu}+\frac{1}{8}\omega_{\mu}^{ab}[\gamma_a,\gamma_b]\approx\partial_{\mu}$, we can calculate the expectation value of the EMT of a single spin-$s$ electron as follows:
\begin{equation} \langle T^{\mu \nu} \rangle\approx\frac{i}{2}\langle k^{s}_j \mid[\bar{\Psi}(\gamma^{\mu}\partial^{\nu}+\gamma^{\nu}\partial^{\mu})\Psi -(\partial^{\mu}\bar{\Psi}\gamma^{\nu}+\partial^{\nu}\bar{\Psi}\gamma^{\mu})\Psi]\mid k^{s}_j\rangle =j^{\mu}k^{\nu}+j^{\nu}k^{\mu} \end{equation} 
in which $\mid k^{s}_j\rangle \equiv \sqrt{2k_0} c_{{k_j}}^{s+} \mid 0 \rangle$ is the single spin-$s$ electron's state with wave vector $k_j$ measured in the LMC, and the electron's current $j^{\mu}$ is given by
\begin{equation}
\begin{split}
j^{\mu}=\langle k^{s}_j \mid\bar{\Psi}\gamma^{\mu}\Psi\mid k^{s}_j\rangle=B_k^{2}\bar{u}^{s}_{k_j}\gamma^\mu u^{s}_{k_j}=-2B_k^{2}k^\mu
\end{split}
\end{equation} 
 which satisfies the continuity equation $\nabla_{\mu}j^{\mu}=0$. Therefore, by plugging Eq. (4.11) into Eq. (4.10), the geodesic equation is reproduced. It is worth noting that our result is different from Ref.~\cite{alsing2009spin}, which shows that spin-1/2 particles exhibit non-geodesic motions caused by the spin connection terms $w_{\mu}^{ab}$, since the spin connections are absent in our calculations with WGF applied.

\section{Particle interactions in curved space-time} 
In previous sections, our analysis reveals that the conventional interpretation of particles as eigen-functions of the field equation in curved spacetime is problematic to represent actual physical particles. Consequently, the Feynman propagator, traditionally calculated as the inverse of the second-order differential operator of the field equation, cannot accurately describe particle interactions in curved spacetime. This limitation underscores the need for an alternative framework that correctly captures the dynamics of particles in such environments.

To address this, we propose using the wave solution provided in Eq. (2.7), which more accurately represents real physical particles in curved spacetime. These solutions align with the expected geodesic motion of particles and offer a foundation upon which a more accurate theory of particle interactions in curved spacetime can be developed. By shifting away from the reliance on eigenfunctions and Feynman propagators in their conventional form, this approach provides a more robust pathway for integrating QFT with the curvature of spacetime, thereby advancing our understanding of particle interactions under the influence of gravity.

In this section, we explore the modifications to Feynman diagrams in Yukawa $\lambda \bar{\Psi} \Phi \Psi$ theory within WSST. For scalar fields in WSST, the non-interacting Hamiltonian of the scalar field is expressed as: 

\begin{equation}
H_0  =\int \frac{d^{3}k_j}{(2\pi)^3}(\omega_{k_j} a_{k_j}^{+}a_{k_j})+const.
\end{equation} 
where $\omega_{k_j}$  is the energy of the $k_j$-th mode, which remains constant if the metric tensor is time-independent.
It is important to note that the states  $\mid \vec{x} \rangle_{t} \equiv  \int \frac{d^{3}{k_i}}{(2\pi)^3}\phi^{\ast}_{k}(t,\vec{x})a_{k_i}^{+} \mid 0 \rangle$  form a basis in real space that is non-orthogonal. Specifically, we have: 
\begin{equation}
_{t}\langle \vec{x}'  \mid \vec{x} \rangle_{t} =  \int \frac{d^{3}{k_i}}{(2\pi)^3}B_k(\vec{x})B_k(\vec{x}') e^{-i\int^{(t,\vec{x})}_{(t,\vec{x}')} k_{\mu} d x^{\mu}} \not = \delta ^{3} (\vec{x}-\vec{x}')
\end{equation} 
This result indicates that gravity reshapes the probability distribution in real space, making it non-uniform and causing these states to become non-orthonormal. Consequently, we must account for the effects of spacetime curvature when describing particle interactions and constructing Feynman diagrams in curved spacetimes.  
The Yukawa interaction is defined by

\begin{equation}
H_{int}  =-\lambda\int d^{3} x\sqrt{S}\bar{\Psi} \Phi \Psi
\end{equation} 
where $S$ denotes the determinant of the 3-spatial metric tensor,  and $\lambda$ is the coupling constant. The Feynman propagator of the scalar particle is expressed as: 
\begin{equation}
\begin{split}
D_{F}(x_1,x_2) 
\equiv\langle 0 |T(\Phi(t_1,\vec{x}_1)\Phi(t_2,\vec{x}_2)) | 0\rangle=
\int \frac{d^{4}k_a}{(2\pi)^4}\frac{iB_k(\vec{x}_1)B_k(\vec{x}_2)}{k_{a}^{2}-m^2+i\epsilon}e^{-i\int^{(t_1,\vec{x}_1)}_{(t_2,\vec{x}_2)} k_{\mu} d x^{\mu}}
\end{split}
\end{equation} 
where $\Phi(t,\vec{x})$ is defined as in Eq. (2.9). Similarly, the Feynman propagator for fermions is

\begin{equation}
\begin{split}
S_{F}(x_1,x_2) 
\equiv\langle 0 |T(\Psi(t_1,\vec{x}_1)\bar{\Psi}(t_2,\vec{x}_2)) | 0\rangle=
\int \frac{d^{4}k_a}{(2\pi)^4}\frac{iB_k(\vec{x}_1)B_k(\vec{x}_2)(\gamma^{a}k_a+m)}{k_{a}^{2}-m^2+i\epsilon}e^{-i\int^{(t_1,\vec{x}_1)}_{(t_2,\vec{x}_2)} k_{\mu} d x^{\mu}}
\end{split}
\end{equation} 
where $\Psi(t_1,\vec{x}_1)$ is defined in Eq. (4.8). 
Notably, these propagators include the additional factor $B_k(\vec{x}_1)B_k(\vec{x}_2)$, reflecting a dependence on the position in the gravitational field. This implies that Feynman diagrams are altered by the gravitational field. 

Considering the spatial integration range in the interacting Hamiltonian of Eq. (5.3), careful attention must be given to how this differs between global Minkowski space-time and curved space-time. In the global Minkowski space-time, integration over the entire space (from $-\infty$ to $\infty$) is straightforward due to the uniformity of fields, serving primarily to enforce momentum conservation. However, in curved space-time, fields are no longer uniform and vary spatially, incorporating factors such as $B_k(x)$ and spatially-dependent momentum $k$. 

In realistic particle physics experiments, interactions typically occur within finite spatial volumes, such as a laboratory setting. Therefore, performing an infinite spatial integration, which would encompass the entire universe, becomes inappropriate for describing localized interactions. For example, when considering an atom within our solar system, where the gravitational field is described by WSST, it would be impractical and unnecessary to integrate over the entire solar system to describe electron interactions within that atom. Since the gravitational effects are encoded in the local metric tensor, and thus, the curvature of space-time far from the localized system should not influence the interactions occurring within it. 

In Minkowski space-time, spatial integration naturally leads to momentum conservation, a principle that should remain valid in WGF as well. The conservation laws are expected to emerge similarly when integrating the phase factor $e^{-i\int k_{\mu} d x^{\mu}}$ in curved space-time, reflecting the localized nature of particle interactions. This localized approach ensures that only relevant spatial regions contribute to the interaction dynamics, aligning with the physical realities of confined experimental settings. 

Given that the wave packets of interacting particles are well-localized around $x_1$, this localized system can be denoted as $\mid s\rangle_{in}$, while the system outside from the interaction region of  $\mid s\rangle_{in}$, representing the state of the rest of the universe, is denoted as $\mid s\rangle_{out}$ . Consequently, the spatial integration in Eq. (5.3) can be split into contributions from within and outside the interaction region: 
\begin{equation}
\begin{split}
\int_{-\infty}^{\infty} d^{3} x\bar{\Psi} \Phi \Psi (\mid s\rangle_{in}+\mid s\rangle_{out})
=
\int_{x_1-\Delta V/2} ^{{x_1+\Delta V/2} }d^{3} x\bar{\Psi}(x) \Phi(x) \Psi(x) \mid s\rangle_{in}+
\int_{outside}d^{3} x\bar{\Psi}(x) \Phi(x) \Psi(x) \mid s\rangle_{out}\\
+\int_{x_1-\Delta V/2} ^{{x_1+\Delta V/2} }d^{3} x\bar{\Psi}(x) \Phi(x) \Psi(x) \mid s\rangle_{out}+
\int_{outside}d^{3} x\bar{\Psi}(x) \Phi(x) \Psi(x) \mid s\rangle_{in}
\end{split}
\end{equation} 
where the volume $\Delta V$ encompasses the interaction region. The two terms in the second line vanish, as the operator and the state are in different spatial zones. The second term in the first line, which accounts for the evolution of the rest of the universe, can be neglected. Therefore, the spatial integration over the entire space reduces to the localized volume where the interaction takes place. Consequently, the S-matrix for particle interactions around position $x_1$ simplifies to: 
\begin{equation}
\begin{split}
_{in}\langle s'\mid T[e^{-i\int dt\int d^{3} x\sqrt{-g}\bar{\Psi} \Phi \Psi}] \mid s\rangle_{in}=
{_{in}\langle} s'\mid T[e^{-i\int dt\int_{x_1-\Delta V/2} ^{{x_1+\Delta V/2} }d^{3} x\sqrt{-g}\bar{\Psi} \Phi \Psi}] \mid s\rangle_{in}
\end{split}
\end{equation} 
Since the momenta $k_{\mu}$ vary slowly with position, they can be approximated as constants during integration over $\Delta V$. For example, in atomic electron interactions, the incoming states $\mid k,p \rangle \equiv \sqrt{2\omega_k}\sqrt{2\omega_p}c^{+}_{k}c^{+}_{p}\mid 0 \rangle$ and the outgoing state $\langle k',p'\mid $ are confined within the Bohr radius. Thus, momenta can be treated as constants during spatial integration: 

\begin{align}
 \begin{split}
\int_{x_1}^{x_{2}} k_{j}e^{j}_{\rho}d x^{\rho}=\int_{x_{1}}^{x_{2}}[k_{j}e^{j}_{\rho}\mid_{x_1}+\frac{\partial (k_{j}e^{j}_{\rho})}{\partial x^{v}}\mid_{x_1}(x^{v}-x^{v}_1)+\mathcal{O}((x-x_1)^2)]dx^{\rho}
\approx k_{j}e^{j}_{\rho}(x_2^{\rho}-x_1^{\rho})
 \end{split}
\end{align} 
where the zeroth-order approximation holds since $k_je^{j}_{\rho}\gg \frac{\partial (k_{j}e^{j}_{\rho})}{\partial x^{v}}\mid_{x_1}(x_{2}^{v}-x^{v}_1)$ with $x_{2}^{v}-x^{v}_1\approx a_0$. Hence, integrating over the interaction volume produces a delta function ensuring momentum conservation: 
\begin{align}
 \begin{split}
\int_{x_1-\Delta V/2} ^{{x_1+\Delta V/2} }d^{3} x \sqrt{S} e^{-i\int^{x} k_{j}e^{j}_{\rho}d x'^{\rho}}\approx \int_{x_1-\Delta V/2} ^{{x_1+\Delta V/2} }d^{3} x \sqrt{S} e^{-ik_{j}e^{j}_{\rho}x^{\rho}}\approx (2\pi)^{3}\delta^{3}(k_j)
 \end{split}
\end{align} 
where $k_j$ is evaluated in the LMC at the interaction site $x_1$. Similarly, the slowly varying function $B_k(x)$, present in $\Psi$ and $\Phi$, can be approximated as $B_k(x_1)$ upon integration over $\Delta V$. Therefore, in this analysis, we focus on short-range interactions, treating $B_{k}(\vec{x}_1)\approx B_{k}(\vec{x}_2)$ in Eqs. (5.3, 5.4) and approximating $k_\mu$ as constants during spatial integration. 

Each mode function in Eq. (2.7) represents a free plane-wave particle across the entire curved spacetime; thus, normalization cannot be limited to a small box, as this would make  $B(x)$ dependent on the box’s size only, ignoring the curvature of the space-time outside of the box. Instead, free particles are treated as wave packets with a localized envelope. For instance, a free KG particle can be represented by a wave function: 
\begin{equation}
    \Phi(t,\vec{x}) =\int d^{3} k f(k)  \phi_{k}(t,\vec{x})
\end{equation}
where $  \phi_{k}(t,\vec{x})$ is given by Eq. (2.7). This particle can be created by applying $\int d^3 k f(k) a_{k}^{+}$ to the vacuum, with  $\int \frac{d^3 k}{(2\pi)^3} \mid f(k)\mid^{2} =1$, ensuring normalization of the particle as indicated by Eq. (2.10). If the envelope function $f(k)$ localizes the particle at $x_1$, the $B(x)$ factor within  $  \phi_{k}(t,\vec{x})$ changes from  $B(x_1)$ to  $B(x_2)$ as it travels, allowing us to observe experimental differences due to the variation in $B(x)$  between $x_1$ and $x_2$, while $f(k)$  remains unchanged. 

In summary, since the wave packets of interacting particles are confined to small regions, $B_k(x)$ and $k$ vary slowly within WGF, these functions can be approximated as constants within the interaction range. Consequently, their values are evaluated at the interaction location, and spatial integration over the interaction region enforces momentum conservation. With this framework, we proceed to outline the modified Feynman diagrams. 

\subsection{Feynman Rules in Yukawa  $\lambda \bar{\Psi} \Phi \Psi$ Theory}

The Feynman rules in curved spacetime are modified to account for the influence of the local gravitational field on particle interactions. The key modifications are as follows:

\begin{enumerate}

\item For each scalar propagator,  \feynmandiagram[horizontal=a to b]{a--[scalar] b}; = ~~$\frac{iB^{2}_k(\vec{x})}{k_{a}^{2}-m^2+i\epsilon}$ in which $x$  represents the interaction location;

 \item For each fermion propagator, \feynmandiagram[horizontal=a to b]{a--[fermion] b}; =~~$\frac{iB_k^2(\vec{x})(\gamma^{a}k_a+m)}{k_{a}^{2}-m^2+i\epsilon}$;
 
\item Multiply the function $B_{k}(x)$ for each external line. For instance,  for scalar and fermion fields, we have $\Phi \mid p \rangle =B_k(x)$ and $\Psi \mid p,s \rangle=B_k(x)u^{s}(p)$,  where $s$ denotes the spin direction and $p$ is evaluated in the LMC at the location of interaction.

\end{enumerate}

To calculate the probability amplitude for a given process, sum over all Feynman diagrams that contribute to the process, incorporating the modified rules for propagators and external lines. Each Feynman diagram should include:

\begin{enumerate}
    \item \textbf{Factors for propagators and external lines} as specified.
    \item \textbf{Imposition of energy-momentum conservation} at each vertex.
    \item \textbf{Integration over undetermined internal momenta} with the measure $\int \frac{d^{4}k_a}{(2\pi)^4}$
\end{enumerate}

\subsection{Example: Two-Fermion Scattering Process }
For a two-fermion scattering process at location $x_1$ in WSST, expanded to second order, the probability amplitude is: 
\begin{equation}
    _{x_1}\langle p',k'\mid \frac{-\lambda^{2}}{2} \int     dx^{4}\sqrt{-g} \bar{\Psi} \Phi \Psi \int dy^{4}\sqrt{-g} \bar{\Psi} \Phi \Psi \mid k,p\rangle_{x_1} 
\end{equation}
The feynman diagram for the scattering process is 
\begin{align}
\begin{gathered}
   \feynmandiagram[horizontal=a to b]{i1--[fermion, edge label'=$p$] a -- [fermion,  edge label'=$p'$] i2 , a -- [scalar] b, f1 -- [fermion,  edge label'=$k$] b -- [fermion,  edge label'=$k'$] f2};
\end{gathered}=
    -\frac{\lambda^2}{2}B_{p}(x_1)B_{k}(x_1)u(p)u(k)\frac{B_{p-p'}^2(x_1)}{(p-p')^2-m^2}B_{p'}(x_1)B_{k'}(x_1)\bar{u}(p')\bar{u}(k') 
\end{align}
where all the momenta are evaluated in LMC established at the interaction site.

The resulting probability amplitude thus encapsulates the impact of the curved space-time on the particle interactions, reflected in the internal and external lines through the factors $B_{k}(x)$. These modifications highlight the dependence of interaction dynamics on the local gravitational field, illustrating how traditional Feynman diagrams from flat space-time QFT are adjusted when generalized to curved backgrounds.

We estimate the impact of the factor  $B_k(x)$ by comparing experiments conducted on the Earth's surface with those at spatial infinity, where space-time approximates Minkowski. For an electron with energy comparable to its rest mass, Eq. (3.7) indicates that the magnitude of  $B_k(x)-1$ is approximately $1\times 10^{-9}$. Although this small deviation falls outside of the detectable range of current experimental capabilities, it still offers a promising opportunity for empirical validation using future technology. 

\section{discussion}
In this study, we have employed the LMC framework within QFT-CS to explore the behavior of quantum particles in the presence of WGF.  A significant contribution of our work is the critical evaluation of the conventional view that particles in curved space-time are simply eigenfunctions of free field equations. We contend that this traditional perspective may be potentially misleading, as it might obscure the true nature of particle dynamics and interactions in curved space-time. 

A key insight from our study is that gravitational effects on particle interactions can be encapsulated in a modified phase factor and a normalization function $B_{k}(x)$, which depends on the curvature of space-time. This approach provides a nuanced understanding of interaction probabilities in curved space-time, emphasizing the pivotal role of the metric tensor and the underlying geometry. 

Building upon this framework, we have extended our analysis to particle interactions using Feynman diagrams in curved space-time. For Yukawa $\lambda\bar{\Psi} \Phi \Psi$ theory, we derived the modified Feynman diagrams, revealing how these differ from their flat space-time counterparts. The adjusted Feynman rules incorporate location-dependent factors that reflect the influence of the gravitational field on interaction dynamics. 

In summary, the modifications to Feynman diagrams and the adjustment of Feynman rules in curved space-time offer a comprehensive perspective on how gravitational fields influence quantum particle interactions, which is crucial for accurate theoretical predictions and experimental interpretations where gravitational effects are significant. Future research could build on these findings by investigating more complex gravitational backgrounds or incorporating higher-order corrections. Additionally, empirical studies in regions with varying gravitational fields could further validate our theoretical predictions and refine the understanding of quantum field theories in curved space-time.

\section{acknowledgement}
I would like to express my sincere gratitude to Professor Roberto Casadio for his invaluable insights during the discussions that contributed to the development of this work.

\appendix


\bibliographystyle{unsrt}
\bibliography{reference}

\end{document}